\documentclass[pra,aps,twocolumn,showpacs,superscriptaddress,amsmath,amssymb]{revtex4}
\usepackage{color}
\usepackage{amsmath,graphicx,bm}

\begin{document}
\def\tr{\rm{Tr}}
\def\la{{\langle}}
\def\ra{{\rangle}}
\def\a{{\alpha}}
\def\e{\varepsilon}
\def\q{\quad}
\def\w{\tilde{W}}
\def\t{\tilde{t}}
\def\a{\hat{A}}
\def\h{\hat{H}}
\def\E{\mathcal{E}}
\def\p{\hat{P}}
\def\ros{R_{\mathrm{so}}}
\def\et{{\bm\eta}}
\def\Ps{{\bm\Psi}}
\def\eti{{\bm\eta}^{I}}

\title{Measurement of non-commuting spin components using spin-orbit interaction.}
%
%
\author {D. Sokolovski$^{1,2}$ and E. Ya. Sherman}
\affiliation{Departmento de Qu\'imica-F\'isica, Universidad del Pa\' is Vasco, UPV/EHU, E-48080 Leioa, Spain}
\affiliation{IKERBASQUE, Basque Foundation for Science, E-48011 Bilbao, Spain}

\date{\today}
\begin{abstract}
We propose a possible experiment aimed at 
a joint  measurement of two non-commuting spin 1/2 components and analyze its physical meaning. 
We demonstrate that switching of a strong spin-orbit interaction, e.g., in a solid state or a cold-atom system, 
for a short time interval simulates a simultaneous von Neumann  measurement 
of the operators $\sigma_x$ and $\sigma_y$. With the spin
dynamics mapped onto the quantum coordinate-space motion, 
such an experiment  
determines averages  of $\sigma_x$ and $\sigma_y$ over the duration of the measurement,
however short the latter may be. These time averages, 
unlike the instantaneous values of $\sigma_x$ and $\sigma_y$, may be evaluated simultaneously to an arbitrary accuracy.

\end{abstract}

%
%
\maketitle

{Recent developments in quantum information 
and technology have brought the quantum measurement theory 
(QMT), originally formulated together with the principles of
quantum mechanics (see, e.g. \cite{vN}) into the research focus. 
Continuing progress in experimental techniques 
has made it possible to test the QMT as well as make new, sometimes surprising, predictions \cite{Steinberg}. 
One fundamental problem in the QMT is that of joint measurement of non-commuting variables which, 
according to the uncertainty principle, cannot 
have well-defined values simultaneously. 
An operational approach to the joint measurement of  particle's position and momentum was proposed in the pioneering work of 
Arthurs and Kelly \cite{AK,SHE,SP9}. Recent attempts to extend it to non-commuting spin components can be found in Refs. \cite{SP0,SP4,SP5}. 
Still, important questions concerning the exact nature of the measured quantity, the accuracy to which its value can be determined, and the 
back-action a exerted on the measured system remain unanswered to this day. 
The purpose of this Letter is to answer these questions, crucial for understanding the 
nature of any quantum measurement.
We also suggest an optimal experimental technique for simulating a joint von Neumann 
measurement on a generic spin-$1/2$ system, of interest in quantum information.
For the latter we propose the use of modern techniques developed for controlling spin-orbit  (SO) interactions in solids 
\cite{Karimov03,Zutic04} and for cold atoms in optical 
lattices  \cite{Stanescu08,Liu09,Wang10,Lin11}. The key feature of such systems, 
currently attracting interest for both fundamental and applied reasons (for a review see \cite{Zutic04}), 
is entanglement between 
the translational and the spin (pseudospin) degrees of freedom. Generated once the SO coupling is switched on, 
the entanglement allows the particle play the role of a von Neumann pointer.
Modulation of the SO coupling strength, including switching it on and off on demand, can 
be achieved for electrons in semiconductor structures by applying external bias to the metallic
gates attached to 
the system \cite{Karimov03,Zutic04}.  For cold atoms similar effect can be realized 
with specially designed optical fields \cite{Stanescu08,Liu09,Wang10,Lin11}.}
\newline
With the above in mind, for a particle of mass $M$, we will consider one of the following Hamiltonians 
($\hbar=1$): 
\begin{equation}\label{1}
\h=g(t)(\hat{p}_x\sigma_{\gamma}\pm\hat{p}_y\sigma_{\delta})+\hat{p}^2/2M,
\end{equation}
where $\hat{p}_x$ and $\hat{p}_y$ are the components of two-dimensional momentum, 
$\hat{p}^2\equiv \hat{p}_x^2+ \hat{p}_y^2$ 
and the indices of the Pauli matrices $\sigma$,
are either $\gamma=x, \delta=y$ or $\gamma=y, \delta=x$. 
We assume the SO interaction to be switched on 
for a finite period of time, 
\begin{eqnarray}\label{2}
g(t) = \alpha \quad {\rm for} \quad 0 \le t \le T,\quad g(t)=0 \quad {\rm otherwise},
\end{eqnarray}
where the coupling parameter $\alpha$ varies from 10 cm/s for cold atoms to 10$^{6}$ cm/s for electrons in semiconductors. 
Thus, for $0<t<T$, 
the operator of particle's velocity depends on the orientation of the particle's spin, $\hat{v}_x= \hat{p}_{x}/M +\alpha\sigma _{\gamma }$,
$\hat{v}_y= \hat{p}_{y}/M \pm \alpha \sigma _{\delta }$.
Without loss of generality, in Eq.(\ref{1}) we choose the
SO coupling in the form $g(t)(\hat{\bf p}\cdot{\bm\sigma})$  \cite{graphene}.
Neglecting the kinetic energy (exact condition will be given further in the text) we have the Schroedinger equation
\begin{equation}\label{3}
i\partial_t\Ps(x,y,t) =-i\alpha(\partial_x \sigma_{x}+\partial_y \sigma_{y})\Ps(x,y,t),
\end{equation}
with an initial condition:
\begin{equation}\label{3:1}
\Ps(x,y,0)=G(x,y)\et^{\rm [in]},
\end{equation}
where $\Ps(x,y,t)$ and  $\et^{\rm [in]}$ are two-component spinors. Using translational invariance of the Hamiltonian (\ref{1}) we rewrite (\ref{3}) as
\begin{eqnarray}\label{4}
\Ps(x,y,t) =\int G(x-x',y-y') \et(x',y',t) dx'dy',
\end{eqnarray}
\begin{eqnarray}\label{4a}
\et(x,y,t) =\la x|\la y| \exp(-iHt)|0\ra|0\ra\et^{\rm [in]}
\end{eqnarray}
where, in addition,
$\int  \et(x,y,t) dxdy = \et^{\rm [in]}$. 
We note that Eqs.(\ref{3})-(\ref{4a}) are identical to those describing 
a spin coupled to two von Neumann pointers  \cite{vN} with positions $x$ and $y$, respectively. which attempt to measure 
two non-commuting projections of the spin simultaneously.
  Considering first measurement of a single spin component, say $\sigma_x$, by choosing $H=-i\alpha\partial_x \sigma_{x}$, offers a useful insight.
  Since operators
  $\exp(-\alpha t \partial_i\sigma_i)$ effect translations in the coordinate space, 
   for $\et^{\rm [in]}$ not an eigenstate of $\sigma_x$, $\Ps(x,t)$ is split into two components which travel in opposite directions with speed $\alpha$. In this way one is able to measure 
  $\sigma_x$ to an accuracy determined by the coordinate width of the initial pointer state.  
Since $\sigma_x$ and $\sigma_y$ do not commute, in a simultaneous measurement the particle which plays the role of the pointer cannot
acquire a well defined velocity.
To study its motion, we slice the time interval $[0,T]$ into $L$ 
subintervals $\varepsilon=T/L$, send $L$ to infinity and apply the Lie-Trotter product formula \cite{TROT} to write
$\exp[-\alpha T(\partial_x\sigma_{x}+\partial_y\sigma_{y})]
=[\exp(-\alpha\varepsilon\partial_x \sigma_{x})$
$\exp(- \alpha\varepsilon\partial_y \sigma_{y})]^L$.
Using the spectral representation for each Cartesian component,  
$\exp(-\alpha\varepsilon\partial_i \sigma_{i})=\sum_{m=\pm1}|m\ra_i \exp(-m\alpha\varepsilon\partial_i)_i \la m|$
where $\sigma_i|m\ra_i=m|m\ra_i$ and $i=x,y$, one readily sees that a pointer undergoes a virtual random walk on 
a lattice $x(j_{x})=j_{x}\alpha \varepsilon$, $y(j_{y})=j_{y}\alpha \varepsilon$, $j_{x},j_{y}=\ldots-1,0,1,\ldots$ reminiscent of Feynman's checkerboard for a Dirac electron  \cite{Feyn}.
\begin{figure}[tb]
\includegraphics*[width=4.7cm, angle=-0]{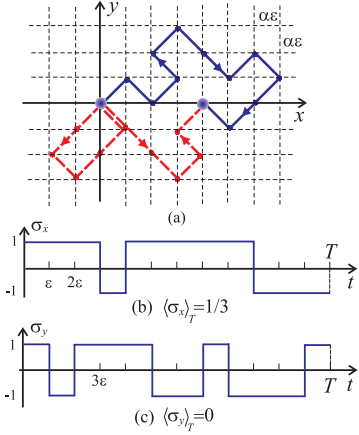}
\caption{(color online) (a) two possible particle's virtual paths in the $xy$-plane
leading to the same final states 
(b) the solid-line path in the $\sigma_x$-subspace with $\la \sigma_x\ra_T=1/3$,
(c) same path in the $\sigma_y$-subspace with $\la \sigma_y\ra_T=0$.
}
\label{fig:FIG1}
\end{figure}
In every time step the particle  moves forwards or backwards along the $x$- and $y$-axes.
Its final position is determined by the differences, $\Delta n_x$ and $\Delta n_y$, 
between the numbers of forward and backward steps taken in each direction or, more precisely, by the interference between all spacial paths sharing the same $\Delta n_x$, and $\Delta n_y$
(see Fig.\ref{fig:FIG1}(a)). Next we assign values $m({l})=\pm 1$, ${l}=1,\ldots,L$ to $\sigma_x$ in each of the subinterval steps, 
write 
$\alpha\varepsilon\Delta n_x=\alpha \varepsilon\sum_{l=1}^{L}m({l})$ and do the same 
for $\sigma_y$.  We note that finding the pointer at a location $(x,y)$ one also determines
{\it time averages} of the spin components, $\la \sigma_x\ra_T$ 
and $\la \sigma_y\ra_T$,
defined for the spin-space Feynman paths  (Figs.\ref{fig:FIG1}(b) and \ref{fig:FIG1}(c)), 
\begin{eqnarray}\label{6a}
\la \sigma_{x,y}\ra_T\equiv T^{-1}\int_0^T \sigma_{x,y}(t) dt,
\end{eqnarray} to an accuracy determined by the position spread of the initial state $G(x,y)$. We note further that a particle initially localized precisely  at the origin can advance along the $x$-axis at most by $\alpha T$ provided all $L$ steps are taken in the positive $x$-direction. 
Since the Hamiltonian in (\ref{3}) is invariant under rotations of 
the coordinate axes, this also implies
that the particle would never leave the 'allowed' circle $r\equiv (x^2+y^2)^{1/2}\le \ros$,
where we introduced the SO coupling determined radius $\ros\equiv\alpha T$.
\newline
To study the distribution of $\la \sigma_{x}\ra_T$ and $\la \sigma_{y}\ra_T$ in detail we return to Eqs.(\ref{3}), where we choose  $G(x,y)$ to be a symmetric 
Gaussian of a width $r_0$ centered at the origin
\begin{eqnarray}\label{6}
G(x,y) = \frac{\sqrt{2}}{\sqrt{\pi}r_0}\exp(-r^2/r_0^2).
\end{eqnarray}  
It is convenient to define four amplitudes $U_{\mu\nu}(x,y,T)$, $\mu,\nu=1,2$ 
for a particle initially at the origin and
with spin initially polarized along ($\nu=1$) or against ($\nu=2$) the $z$-axis, 
to be found at $t=T$ 
at a location $(x,y)$, polarized along ($\mu=1$) or against ($\mu=2$) the $z$-axis.
Performing a Fourier transform of Eq.(\ref{3}) with respect to $x$ and $y$,
we find, in the cylindrical 
coordinates, that $U_{\mu\nu}(r,\theta,T)$ is a Hermitian matrix whose elements are \cite{zitter,inglot}
\begin{eqnarray}\label{7}
&&U_{11}(r,T)=U_{22}(r,T)=\sqrt{2\pi}r_{0}\times\\
\nonumber
&&\int_{0}^{\infty
}\exp \left(-k^{2}r_{0}^{2}/4\right) \cos \left( R_{{\mathrm {so}}}k\right)
J_{0}(kr)\frac{kdk}{2\pi },\\
\nonumber
&&U_{12}(r,\theta,T)=U^{*}_{21}(r,\theta,T)=\sqrt{2\pi}e^{-i\theta}r_{0}\times\\
\nonumber
&&\int_{0}^{\infty }\exp \left( -k^{2}r_{0}^{2}/4\right) \sin \left( R_{%
{\mathrm {so}}}k\right) J_{1}\left( kr\right) \frac{kdk}{2\pi },
\end{eqnarray}
where $J_n(z)$ is the Bessel function of the first kind of order $n$ and
$\theta$ is the angle the vector $(x,y)$ makes with the $x$-axis.
We are interested in an accurate measurement, where the maximum shift of the particle (pointer),
$R_{{\mathrm {so}}}$, is much greater than the width of the initial Gaussian
\begin{eqnarray}\label{8}
r_0/R_{{\mathrm {so}}}\ll 1,
\end{eqnarray}
in which case main contributions to the integrals in (\ref{7}) come from the region where $k R_{{\mathrm {so}}}\gg 1$,
and the condition for neglecting kinetic energy reads
$T\ll r_0^{2}{M}$. 

To evaluate possible switching times $T$  for electrons in semiconductors, we use recent estimates \cite{Karimov03,Zutic04} $r_{0}\sim 10^{-5}$ cm, 
$\alpha\sim 10^{6}$ cm/s, and the effective mass $M\sim 10^{-28}$ g, thus obtaining $T\sim 10^{-11}$s,
meaning that $r_0/R_{{\mathrm {so}}}$ cannot be less than 0.1. 
For cold atoms  with \cite{Stanescu08,Liu09} $r_{0}\sim 10^{-4}$ cm, $\alpha\sim 10$ cm/s,
and atomic mass of $M\sim 10^{-22}$ g, we obtain a broader range of durations, 
$10^{-5}\le T \le 10^{-3}$ s,
suitable for an accurate simulation of a joint von Neumann measurement of the two spin components.

\begin{figure}[tb]
\includegraphics*[width=5.2cm, angle=-0]{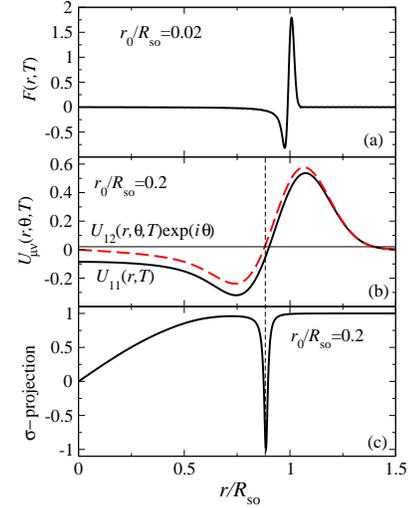}
\caption{(color online) a) $F(r,T)$ in Eq.(\ref{9}) for $r_0/\ros=0.02$.  Its approximation in Eq.(\ref{12}) 
coincides with $F(r,T)$ with the graphical accuracy. This small $r_0/\ros$ ratio can be realized in 
cold atomic gases;
b) $U_{11}(r,T)$ (solid line) and $U_{12}(r,\theta,T)$ (dashed line) for $r_0/\ros=0.2$ (realizable
in semiconductor spintronics) calculated with exact Eq.(\ref{7}); c) the spin projection $\bar{\sigma}_{{\bm \bar{v}}}$  as a function of the coordinate. 
}
\label{fig:FIG2}
\end{figure}

\begin{figure}[tb]
\includegraphics*[width=7cm, angle=-0]{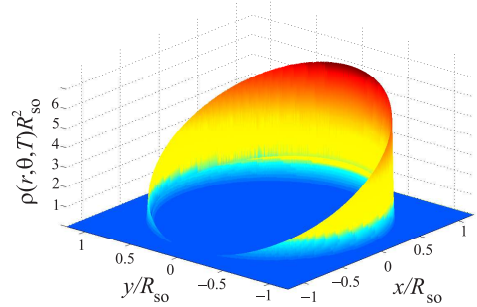}
\caption{(color online) Probability distribution $\rho(r,\theta,T)$ for a spin initially polarized along the $x$-axis,
$\et^{\rm [in]}=(1,1)^{T}/\sqrt{2}$ and $r_0/\ros=0.01$.
}
\label{fig:FIG3}
\end{figure}

\begin{figure}[tb]
\includegraphics*[width=4.0cm, angle=-0]{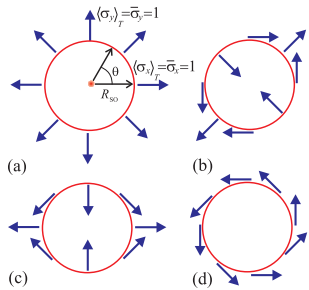}
\caption{(color online) Spin configuration for different forms of the SO
coupling Hamiltonians: (a) $\hat{p}_{x}\sigma_{x}+\hat{p}_{y}\sigma_{y}$,
(b) $\hat{p}_{x}\sigma_{y}+\hat{p}_{y}\sigma_{x}$, (c) $\hat{p}_{x}\sigma_{x}-\hat{p}_{y}\sigma_{y}$,
and (d) $\hat{p}_{x}\sigma_{y}-\hat{p}_{y}\sigma_{x}$.}
\label{fig:FIG4}
\end{figure}

Replacing the Bessel functions by their large argument asymptotes, \cite{ABRAM}
$J_n(z)\sim (2/\pi z)^{1/2}\cos(z-n\pi/2-\pi/4)$ and neglecting oscillatory terms in the integrand yields
\begin{eqnarray}\label{9}
&&\hat{U}(r,\theta,T)\approx F(r,T) \left[ 
\begin{array}{ll}
1 &\exp(-i\theta) \\ 
\exp(i\theta) & 1
\end{array}\right],\\
\nonumber
&&F(r,T)=\frac{r_{0}}{2\pi\sqrt{\ros}}\times\\
\nonumber
&&\int_{0}^{\infty}\exp \left( -k^{2}r_{0}^{2}/4\right) \cos[ (r-R_{{\mathrm {so}}})k+\pi/4]
k^{1/2}dk. 
\end{eqnarray}
For a small $r_0$,  the radial function $F(r,T)$ shown in Fig.\ref{fig:FIG2}(a) has a maximum and a minimum close
to $r=\ros$, rapidly decreases for $r>\ros$, and exhibits a somewhat slower decay for $r<\ros$.
This behavior is understood by noting first that for $r_0=0$
the integral in  the first of Eqs.(\ref{7}) can
be calculated exactly  
in the Cartesian coordinates
 ($k\equiv(k_x^2+k_y^2)^{1/2}$),
\begin{eqnarray}\label{11}
&&\int_{0}^{\infty} \cos \left( R_{{\mathrm {so}}}k\right)
J_{0}(kr)\frac{kdk}{2\pi }\\
&&=\frac{\partial }{\partial R_{%
{\mathrm {so}}}}\int_{-\infty }^{\infty }\frac{dk_{x}dk_{y}}{\left( 2\pi
\right) ^{2}}\frac{\sin (kR_{{\mathrm {so}}})}{k}\exp (ik_{x}x+ik_{y}y). 
\nonumber
\end{eqnarray}
The Fourier transform of $\sin(kz)/k$ is known \cite{FOLL} to be $(2\pi)^{-1}(z^2-r^2)^{-1/2}\chi_z(r)$
where the characteristic function $\chi_z(r)=1$ for $r\le z$ and $0$ otherwise. For a finite $r_0$, convolution 
(cf. Eq.(\ref{4})) of (\ref{11}) with the Gaussian (\ref{6}) yields
an alternative highly accurate form for $F(r,T)$ 
\begin{eqnarray}\label{12}
F(r,T) \approx -\frac{1}{\sqrt{\ros}}\frac{\partial}{\partial r}\int_0^{\ros}
\hspace{-0.2cm}r'\frac{\exp[-(r-r')^2/r_0^2]}{4\sqrt{\ros-r'}}dr', 
\end{eqnarray}
and, therefore for $U_{11}(r,T)$. 
From Eq.(\ref{12}) it is readily seen that for small $r_0$, the integral in Eq.(\ref{11}) behaves as $(\ros-r)^{-1/2}$ if
$\ros -r > r_0$, peaks at $r\approx \ros$ when the center of the Gaussian 
coincides with the integrable singularity of $(\ros-r')^{-1/2}$. For $r>\ros$ it rapidly decays as the overlap
of the Gaussian with the interval $[0,\ros]$ decreases. Accordingly,
$F(r,T)$ behaves as $-(\ros-r)^{-3/2}$ for $r<\ros$, passes through a zero at $r\approx \ros$,
and rapidly decays outside the allowed circle $r\le \ros$. Thus, with the help of  (\ref{9}) for the 
wavefunction (\ref{4}) we find (indices $1$ and $2$ are used for the spin projections up and down the
$z$-axis, respectively), 
\begin{eqnarray}\label{13}
\left[\begin{array}{l}
\Psi_1(r,\theta,T) \\ 
\Psi_2(r,\theta,T)
\end{array}\right]=
F(r,T)\left[\begin{array}{l}
\eta^{\rm [in]}_1+\eta^{\rm [in]}_2\exp(-i\theta) \\ 
\eta^{\rm [in]}_1\exp(i\theta)+\eta^{\rm [in]}_2
\end{array}\right],
\end{eqnarray}
\newline
concentrated in a narrow ring of a radius $\ros$ and a width $\approx r_0$.
From Eq.(\ref{13}) we obtain the probability to find the particle at a location $(r, \theta)$,  
\begin{equation}
\label{15}
\rho (r,\theta,T)=|F(r,T)|^{2}\left[1+\left( \mathbf{n}_{\theta }\cdot 
{\bm \sigma }^{\rm [in]}\right)\right]
\end{equation}
where $\mathbf{n}_{\theta }=\left( \cos \theta ,\sin \theta \right) ,$ and 
$\mathbf{\sigma}_{\gamma}^{\rm [in]}=\la\et^{\rm [in]}|\sigma_{\gamma}|\et^{\rm [in]}\ra$
consists of the spin components for the initial state {(see, e.g., \cite{zitter,Govorov,Molnar,Berman} for 
solid-state realizations)}. With the Hamiltonian in (\ref{3}) invariant under rotations in the $xy$-plane, angular dependence
in Eq.(\ref{15}) comes from the asymmetry of $\et^{\rm [in]}$. In particular, for a spin
whose initial direction is normal to the plane 
i.e., for $\et^{\rm [in]}=(1,0)^{T}$ or $(0,1)^{T}$ the distribution
is isotropic, $\rho(r,\theta,T)=|F(r,T)|^2$. Figure \ref{fig:FIG3} shows $\rho(r,\theta,T)$ for a spin initially directed along the $x$-axis.
\newline
While in an accurate (ideal) measurement position of the particle $(x,y)$ correlates
with the time averages $\la \sigma_{x}\ra_T$ and $\la \sigma_{y}\ra_T$, its relation to the final 
spin orientation is less direct. As seen from the Feynman path analysis, at some point $(x,0)$, the amplitudes 
to have polarizations along and against the $x$-axis, build up from the paths in 
Fig.\ref{fig:FIG1}(a) which arrive at $x$ from the left (e.g., dashed line)  and from the 
right (e.g., solid line), respectively. Since no paths arrive
at $x=\ros$ from the right, the spin at that point (and elsewhere on the circle shown in Fig.\ref{fig:FIG4}(a)) is
always pointing outwards,
\begin{eqnarray}\label{14}
 \bar{\sigma}_{x}(r,\theta,T)=\cos\theta, \q  \bar{\sigma}_{y}(r,\theta,T)=\sin\theta, 
 \end{eqnarray}
 where $\bar{\sigma}_{\gamma}(r,\theta,T)\equiv \la\Ps|\sigma_{\gamma}|\Ps\ra/(|\Psi_1|^2+|\Psi_2|^2)$.
This is also true for any initial spin state $\et^{\rm [in]}$ which, 
 since the matrix in Eq.(\ref{9}) is singular, cannot be reconstructed from the spinor in the r.h.s. of Eq.(\ref{13}).
Although in an ideal measurement the mean pointer's velocity, $\bar{v}_{i}\equiv r_{i}/T$,  $\la \sigma_{i}\ra_T$ and $\bar{\sigma}_{i},$ are simply related, $\bar{v}_{i}=\alpha \la \sigma_{i}\ra_T=
\alpha \bar{\sigma}_{i}$, this is no longer true for a less accurate measurement.
For $r_0\sim \ros$, the final spin state $\Ps(x,y,T)$  is a superposition of all $\et(x',y',T)$ which fit under the Gaussian  in Eq.(\ref{4}) centered at $(x,y)$ and the correlation between $\bar{v}_{x,y}$ and $\bar{\sigma}_{x,y}$ may be lost due to interference. Figure \ref{fig:FIG2}(c) shows 
the projection of the final spin onto the particle's mean velocity, $\bar{\sigma}_{{\bm \bar{v}}}$.
The narrow range of $r$'s where the particle arrives with its spin reversed with respect to
 its mean velocity,  $\bar{\sigma}_{{\bm \bar{v}}}\approx -1$, is a result of such an interference. The condition 
for $\bar{\sigma}_{{\bm \bar{v}}}$ to equal  $ -1$ is $U_{11}(r,T)=-U_{12}(r,\theta,T)\exp(i\theta)$ so that the  resonance-like feature  in Fig.\ref{fig:FIG2}(c)  persists for 
finite values of $r_0/\ros$ and vanishes for $r_0/\ros\rightarrow 0$ when 
the two curves in Fig.\ref{fig:FIG2}(b) effectively coincide (cf. Eq.(\ref{9})).
\newline
Our analysis is extended to other Hamiltonians in Eq.(\ref{1}) by replacing the angle $\theta$ in 
Eq.(\ref{9}) by $-\theta$ or $\pi/2\pm \theta$, as appropriate. Figure \ref{fig:FIG4} shows,
for $r_0/\ros \rightarrow 0$, the spinor field at $t=T$ for each of the four possible cases. 
\newline
In summary, we have shown that switching a strong SO coupling over a short time $T$,
in a solid state or cold-atom system, simulates a simultaneous von Neumann measurement of two non-commuting spin components. 
In this case, the particle plays the role of a pointer which correlates its position, $(x,y)$,  with the 
time average of the corresponding
spin components, $\la\sigma_x\ra_T$ and $\la\sigma_y\ra_T$ evaluated along Feynman paths defined for the 
two spin variables (Fig.\ref{fig:FIG1}(b) and \ref{fig:FIG1}(c)).
There are infinitely many trajectories which share the same values of $\la\sigma_x\ra_T$ and $\la\sigma_y\ra_T$, 
leading to the same pointer position. 
An accurate measurement reveals that the time averages obey the sum rule
$\la\sigma_x\ra_T^2+\la\sigma_y\ra_T^2=1$, with angular anisotropy of the distribution
determined by the anisotropy of the initial spin state.
Importantly, $\la\sigma_x\ra_T$ and $\la\sigma_y\ra_T$
whose values can be  determined in a generic
joint von Neumann measurement to an arbitrary accuracy,  do not represent 
`instantaneous' quantum mechanical expectation values of the operators 
$\sigma_x$ and $\sigma_y$
and do not reduce to these, no matter how short the measurement is.
Indeed, the highly irregular fractal-like Feynman paths shown in Fig.\ref{fig:FIG1}
have no intrinsic time scale of their own. Thus, even in the impulsive limit  $T\rightarrow 0$,
$\alpha T={\rm const}$ one does not attain unique instantaneous values of the two spin components. 
No matter how short $T$ is, all paths (cf. Fig.\ref{fig:FIG1}(a)) contribute to the transition
amplitude (\ref{4}). This is one particular way of stating that non-commuting quantities $\sigma_x$ and $\sigma_y$ 
cannot both have well defined values at the same time.
\newline
We acknowledge support of the University of Basque Country UPV/EHU grant GIU07/40,
MCI of Spain grant FIS2009-12773-C02-01, and "Grupos Consolidados UPV/EHU 
del Gobierno Vasco" grant IT-472-10. We are grateful to A. Eiguren for valuable
discussions.

\end{document}